\documentclass[prl,aps,showpacs,twocolumn]{revtex4}

\begin{document}

\title{
Coherent radiation from neutral 
 molecules moving above a grating
} \author{Alexey Belyanin$^{1,2}$, Federico Capasso$^3$, 
Vitaly  Kocharovsky$^{1,2}$, and Vladimir Kocharovsky$^2$
}
\affiliation{$^1$Physics Department and Institute for Quantum 
Studies, \\
Texas A\&M University, College Station, TX 77843-4242
 \\
$^2$ Institute of
Applied Physics, Russian Academy of Science, \\ 46 Ulyanov 
Street,
603600 Nizhny Novgorod, Russia \\
$^3$Bell Laboratories, Lucent Technologies, \\ 600 Mountain 
Avenue,
Murray Hill, NJ07974, USA 
}

\date{\today}

\begin{abstract}

  We predict and study the quantum-electrodynamical effect of parametric 
self-induced  
  excitation of a  molecule moving above the 
  dielectric or conducting medium with periodic grating. In this case 
  the radiation reaction force  
  modulates the molecular transition frequency which results in a 
  parametric instability of dipole oscillations even from the level 
  of quantum or thermal fluctuations.  The present mechanism of 
instability of  electrically neutral molecules is different from that 
of  the well-known Smith-Purcell and transition radiation in which a 
moving charge and its oscillating image create an oscillating dipole. 
 We show that parametrically excited molecular bunches can 
produce an easily detectable coherent radiation flux of up 
to a microwatt. 

\end{abstract} 


\maketitle

\noindent 
{\bf Introduction.}
 The presence of conducting 
or dielectric surfaces near an atom or a  molecule modifies its   
properties in a fundamental way, changing its radiation and back reaction 
as well as the electromagnetic  vacuum  
fluctuations, and giving rise to a number of interesting effects; see,  e.~g., 
  \cite{Dr74,Ch78,Hi91,Mo69,Ar88,Kh91,Or94,Ch95,Me92,Ya92}. In 
  particular, the atomic energy levels and the radiative 
 decay rate are changed. A boundary can alter also the dynamics of 
 atomic or molecular dipole oscillations. Here we describe a new 
 effect of the latter type.

The calculations of the level shifts for a molecule 
near a perfectly conducting wall can be traced back to the papers
\cite{Mo69,Ca48,Ba70}. According to subsequent
works \cite{Mi82,Da82,Me90,Hi91a}
(and       in
accordance        with fluctuation-dissipation theorem),   this
effect   is   due   to  the modification of both electromagnetic  vacuum
fluctuations  and    radiation reaction. Note, however, that in the near
zone $R \ll \lambda_0/\sqrt{\varepsilon_1}$ the influence  of  a boundary 
between two media with dielectric constants $\varepsilon_1$ and $\varepsilon_2$ on 
a radiating dipole is dominated by the radiation reaction force, while the contribution from vacuum fluctuations of the electromagnetic field is negligible. This result has been obtained in \cite{Me90} using the ideas of separating the self-action and vacuum-fluctuation terms developed in \cite{Da82}; see \cite{Hi91} for the review. The leading term in the self-action force  
 is proportional to a large factor  
$(\lambda_0/R)^3$ and can be interpreted as a non-retarded 
London-van-der-Waals interaction with an instantaneous image dipole. Here $\lambda_0 = 2 \pi c/\omega_0$ is the vacuum 
 wavelength  of a given dipole transition with frequency $\omega_0$, 
$\varepsilon_1$ is the dielectric constant of the medium in which an atom is 
located.  This large factor originates from the near field 
\begin{equation} \label{4} {\bf E}_{\bot, \parallel } = - \frac{(\mp 3 - 1) {\bf 
p'_{\bot, \parallel}}}{16 R^3\varepsilon_1}, \quad  {\bf p'_{\bot, \parallel}} = \mp \frac{{\bf 
   p}_{\bot, \parallel}(\varepsilon_1 - \varepsilon_2)}{\varepsilon_1 + \varepsilon_2}, \end{equation}
created by a high-frequency image dipole ${\bf p'}$ at the position of a real dipole
${\bf p}$.
Here and below upper and  lower  signs  correspond  to  the
dipoles oriented  perpendicular $(\bot)$ and parallel $(\parallel)$ to
the boundary.

 There is a number of reasons why the dynamics of molecular radiative 
 transitions is modified in the presence of boundaries. The effects 
 that were identified and observed include 
     (i) changes   in   the   spectral   density   of  radiated  modes
\cite{Hi91,Or94,Ch95,Me92,Ba87}, 
 (ii) location of an atom in the nodes or maxima of resonant modes 
 or in a non-transparent medium, e.~g., in a medium with  
 negative dielectric  constant or in a photonic band structure, when 
 the transition frequency  is inside the Bragg gap 
 \cite{By72,Sa94,Su95}, (iii) phase shift of the near field  of a 
 dipole due to dissipation in the neighboring medium 
\cite{bkk95,bkk96}.

 Additional possibilities to affect the molecular dynamics arise when 
 the  dipoles are moving near the interface of two media or inside a 
 medium. A well-known example is Cherenkov radiation of an 
 oscillating dipole moving in an anisotropic medium under the 
 conditions of anomalous Doppler effect which can lead to instability 
 of the dipole oscillations \cite{gin}. This instability is, however, 
 very difficult to realize since radiation in the directions 
 corresponding to the normal Doppler effect usually dominates.  

 In this paper we investigate a new mechanism of boundary-induced 
 excitation of molecular transitions which is realized when a 
 molecule is moving close to the dielectric or conducting medium with 
 periodic grating. In this case the radiation reaction force acting 
 on an oscillating dipole moment of a given transition is mainly due to 
the time-dependent London-van-der-Waals interaction experienced by the molecule moving above a grating. It is 
 proportional to the instantaneous value of the dipole moment with a 
 proportionality coefficient being a periodic function of time. Its  
 modulation frequency $\nu$ is equal to the velocity 
 of a molecule divided by the spatial period of a grating.  Such a 
 modulated radiation reaction force represents a 
 periodic perturbation of the transition frequency $\omega_0$ 
 (eigenfrequency of dipole oscillations) which can drive a 
 parametric instability under the conditions of Mathieu's 
 resonance: $\nu = 2 \omega_0/N$, where $N$ is $1,2,3 \dots$ In 
 other words, the instability is due to the periodically modulated 
 radiation reaction force which gives rise to excitation of dipole 
 oscillations even from the level of quantum or thermal fluctuations.  
 
 If we have a beam of molecular dipoles, one 
 can expect generation of measurable coherent radiation flux. 

 We emphasize that this situation is different from the well-known 
 Smith-Purcell effect \cite{sp} in which a charge moving above a 
 periodic grating creates an oscillating dipole due to its 
 oscillating image. It is also different from the transition radiation of a charge traversing a periodic dielectric stack \cite{tr}.  In the present case, it is the electrically 
neutral atom or molecule that gets parametrically excited. 

 To find the conditions for parametric instability, we calculate the 
 back reaction force acting on a dipole  oscillator  with  a  
high-frequency  dipole moment ${\bf p}(t)$ 
 which is moving with a constant velocity 
$v$ in a medium with real dielectric constant $\varepsilon_1$ near the 
boundary of a medium with complex dielectric constant $\varepsilon_2$.  
Precisely, we calculate the electric field $ {\bf E}(t)$ 
 created by a 
high-frequency dipole at its position $z = R$, $x = vt$, $y = 0$. We 
assume that the medium 2 has a periodic grating with period $L$ in 
the direction $x$ along the dipole velocity, so that the distance $R$ 
between the dipole and the boundary is actually a periodic function 
of time with period $T = L/v$. 

 We assume that the distance $R$ to the boundary is much less than 
 the grating period, so that the radiation reaction field at any 
 moment of time $t$ is approximately equal to the one for the dipole 
 at a distance $z(t)$ above an infinite plane boundary. This is of 
 course the most interesting limit. In the opposite case, 
when $R \gg L$, the dipole will ``feel'' simultaneously many grating 
periods and the effect of grating will be averaged almost to zero.  

The problem of the radiation reaction field acting  on a dipole above a 
plane boundary was considered many times. 
We will use here the results of our papers \cite{bkk95,bkk96}. 
According to these works, the radiation reaction field can be 
represented as a sum of a free-space term 
     $ {\bf E}_{\rm free}(t)       =       
 2{\bf      \stackrel{\cdots}{p}}(t) \sqrt{\varepsilon_1}/3c^3$ and 
a boundary-induced term ${\bf E}_b$ that can be very complicated. 

     If the  medium  2  is  a  perfect conductor,  $\varepsilon_2 \rightarrow
i\infty$, then the contribution of a boundary  is  
equivalent to  the  field  of an image dipole ${\bf p}'_{\|} = - {\bf
p}_{\|}$ or
${\bf p}'_{\bot} =  {\bf
p}_{\bot}$, retarded by the time $ t_1 = 2R\sqrt{\varepsilon_1}/c$:
     \begin{eqnarray} \label{6} {\bf E}_{b \bot, \parallel}(t)
&=& \frac{(1 \mp 1)\ddot{{\bf p}}_{\bot, \parallel}(t - t_1)}{4Rc^2 } +
\frac{\dot{{\bf p}}_{\bot, \parallel}(t  -
t_1)}{(3 \mp 1) R^2 c \sqrt{\varepsilon_1}} \nonumber \\ 
&+& \frac{{\bf p}_{\bot, \parallel}(t - t_1)}{2(3 \mp 1)R^3 \varepsilon_1}. 
\end{eqnarray} This result  was  obtained  in  many  works,  
including  the full quantum-electrodynamical treatment of the 
problem; see Refs.~[1-5,13,15,17,19,22,23]. 

 To illustrate the physical mechanism of an instability we will 
 assume in this paper that the medium 2 is a perfect conductor and 
 use the above expression (\ref{6}) for the field. The case of an 
 arbitrary medium is qualitatively similar and can be analyzed by 
 using general expressions for the field obtained in 
\cite{bkk95,bkk96}. 
\\ 

\noindent 
{\bf Parametric instability of a harmonic dipole oscillator.} 
This case is relevant for dipole transitions  in a system with many 
quasiequidistant levels, for example, vibrational transitions in a 
molecule or Rydberg transitions in an excited atom. 
Such a system has essentially classical dynamics. 
The equation for free dipole
oscillations of a classical harmonic oscillator with a charge $e$,
mass $m$, and frequency
$\omega_0$ 
  takes the form
\begin{equation}
\label{7} 
d^2{\bf p}/dt^2 + \omega_0^2{\bf p} = (e^2/m) ({\bf E}_{\rm free}+{\bf E}_b). 
 \end{equation} 

 In the free space, this equation describes an oscillator decaying 
 with radiative rate $ \gamma = 2 e^2 \omega_0^2 
\sqrt{\varepsilon_1}/(3 m c^3)$.  
 
In the presence of a boundary, 
 let us consider for definiteness the dipole oriented perpendicular 
 to the interface between two media. The result for a parallel 
 dipole is the same apart from a different numerical 
 coefficient. Then, expanding the electric field ${\bf E}_{b \bot}$ 
 in Eq.~(\ref{6}) in powers of a small parameter $R/\lambda_0$, we 
 obtain the following equation: \begin{equation} \label{8} 
 \ddot{p} + 2 \gamma \dot{p} + \left(\omega_0^2 - 
 \frac{e^2}{4mR^3\varepsilon_1}\right) p = 0.  \end{equation} 

The factor $2\gamma$ describes the well-known result that a 
 perpendicular dipole above the metal surface decays with a rate 
 twice that in a free space. Our  main interest here is in the 
 frequency shift factor which is proportional to $1/R^3$ and can have 
 much larger value than radiative decay proportional to 
 $1/\lambda^3$. Note that for an arbitrary medium 2 with  complex 
 dielectric constant $\varepsilon_2$ this frequency shift has also imaginary 
 part $\propto \varepsilon_2''/R^3$ that can be larger than $\gamma$.  This 
can lead to strong modification of the spontaneous emission rate and even 
reversal of the direction of radiative transitions 
\cite{bkk95,bkk96}. 

 For an ideal conductor the frequency shift is real, and the only 
 possible mechanism of instability is parametric excitation of dipole 
 oscillation due to periodic resonance modulation of this term. This 
 can be achieved by, e.g., modulating the distance $R$ to the 
 boundary. Such a modulation occurs when, for example, the molecule 
 is moving above a periodic grating of period L.  Note that we 
consider here {\it free} dipole oscillations, without any external 
electromagnetic field force. 

 Suppose for simplicity that the grating is sinusoidal. Then the 
 distance $R$ in the expression for the radiation reaction field will 
 be periodically modulated as $$ R = R_0(1+a \cos\nu t), $$ where 
$\nu = 2\pi v/L$. The parametric resonance occurs when 
$\nu = 2\omega_0/N$, 
for an integer $N = 1,2,...$ . 

 If the relative grating amplitude is small, $a \ll 1$, we obtain 
 from Eq.~(\ref{8}) the Mathieu's equation:  \begin{equation} 
\label{9} 
\ddot{p} + 2 \gamma \dot{p} + \omega_0^2(1 +A \cos \nu t) p = 0,  
 \end{equation} 
where 
\begin{equation}
\label{10} 
A = \frac{3 e^2 a}{4 m R_0^3 \omega_0^2 \varepsilon_1}. 
\end{equation} 
The condition for instability on the main resonance $N = 1$ reads 
$ A > 4\gamma/\omega_0$,  
which can be rewritten as 
\begin{equation} \label{cond} 
\frac{9}{256 \pi^3 \varepsilon_1^{3/2}} \left(\frac{\lambda_0}{R_0}\right)^3 a > 1. 
\end{equation} 
 Assuming $a = 0.1$, the numerical factor on the left-hand side of 
 (\ref{cond}) is of order  $10^{-4}$, which requires $\lambda_0/R_0 > 
 20$. Since $R_0$ cannot be reasonably made much smaller than 0.1 
 $\mu$m, we need to use long-wavelength transitions, e.g., 
 vibrational-rotational transitions in molecules.  

In addition, the resonance condition implies $\lambda_0 = 2cL/v$. 
 The value of $L$ is bounded from below:  $L > R_0 \gtrsim 0.1$ 
 $\mu$m. The velocities of neutral molecules achieved in supersonic 
 jets or Laval nozzles are of the order of 1 km/s. In this case 
 $\lambda$ falls into the cm range corresponding to rotational 
 transitions that are better described by a two-level model (see 
 below). However, by accelerating charged molecules and then 
 neutralizing them much greater velocities can be achieved that 
 allow one to employ IR and sub-mm vibrational transitions with 
 quasi-equidistant energy levels. 

Not too close to the threshold, the growth rate of dipole 
oscillations is \begin{equation} \label{gr} \omega'' \simeq \omega_0 A/4 
\simeq 3 e^2 a/(16 m \omega_0R_0^3 \varepsilon_1).  \end{equation} Using 
the correspondence principle we can relate the classical dipole 
amplitude with the matrix element $d$ of a given dipole transition:  
 $e^2/m = 2 \omega_0 d^2/\hbar$. Then Eq.~({\ref{gr}) yields the 
 growth rate \begin{equation} \label{grq} \omega'' = 3 d^2 a/(8 
\hbar R_0^3 \varepsilon_1).  \end{equation} For a typical dipole moment $d 
 \sim 1$ Debye and $R_0 \sim 0.1$ $\mu$m, we get $\omega'' \sim 5\times 
 10^4$ s$^{-1}$, which means that an excitation length $v/\omega''$ is 2 
 cm.  Therefore, the excitation length is short enough to make  
 the effect readily observable in experiment.  
\\

\noindent 
{\bf Instability of a two-level system.}
 An opposite limiting case of a molecular transition is a two-level 
 system. A good example 
 in the low frequency range is rotational transitions in molecules. 
One can obtain the 
modified Bloch equations for a two-level molecule in the vicinity of 
an interface following the usual derivation of 
master equations for a two-level system in a free space. 
The result is presented below, again for a perpendicular dipole: 
\begin{eqnarray} \ddot{\bf p} + 2 \gamma 
\dot{\bf p} + (\omega_0^2 & -& d^2 \omega_0 \Delta n/(2   
\hbar R^3)) {\bf p} \nonumber \\ &  =& 2 \Delta n d^2 
\omega_0 \hbar^{-1} {\bf E}_{{\rm ext}},  \label{20} \\ \label{21} 
d(\Delta n)/dt + \gamma (\Delta n - \Delta n_p) &=& - 
(2/\hbar \omega_0) {\bf E}_{{\rm ext}} \dot{{\bf p}}.  
\end{eqnarray} Here $\Delta n = n_1 - n_2$ is volume density of the 
population difference between two states, $\Delta n_p$ is the 
population difference supported by an incoherent pumping, ${\bf 
E}_{{\rm ext}}$ is an external (classical) field which does not 
contain any back-reaction field. 

 In the absence of an external field we obtain the growth rate of the 
 parametric instability, 
 \begin{equation} \label{grq2} \omega'' = \frac{3 d^2 a \Delta n}{8 
\hbar R_0^3 \varepsilon_1},  \end{equation} which is different from 
 Eq.~(\ref{grq}) only by a factor $\Delta n$. This factor can be 
 rather small for the rotational transitions at room temperature. 
   This may require low temperatures for more efficient excitation. 
 To reach $\Delta n \sim 1/2$ for $\omega_0 = 10^{11}$ s$^{-1}$, 
 liquid helium temperatures are required.  

 The value of the growth rate can be enhanced even further by 
 launching molecules closer to the surface or decreasing the 
 dielectric constant $\varepsilon_1$. It is difficult to send molecules 
 closer than 0.1 $\mu$m to the surface. At the same time, the 
 dielectric constant can be drawn to very small values by creating a 
 rarefied background plasma with plasma frequency close to the 
 frequency $\omega_0$ of dipole oscillations. For $\omega_0 = 10^{11}$ 
 s$^{-1}$ the density of ionized gas should be around $10^{12}$ 
 cm$^{-3}$. The plasma of such density can be very easily produced in 
 a gas discharge. Collisions in such a rarefied plasma are 
 unimportant, and the residual value of $\varepsilon_1$ will be defined by 
 density  close to the grating surface, where the double electric 
 layer is formed.  Experiments indicate the possibility to decrease 
 $\varepsilon_1$ down to $10^{-4}$. 

 Such an enhancement of the parametric instability in a rarefied 
 plasma is interesting by itself. The experimental realization of 
 this effect is facilitated by the fact that fluctuations in $\varepsilon_1$ 
 are not important since the quantity $\varepsilon_1$ in all expressions for 
 the growth rate is actually the value averaged over a very large 
 spatial scale $\lambda_1 = \lambda_0/\sqrt{\varepsilon_1}$. 
\\ 

\noindent 
{\bf Possible experiments.} 
There is a number of ways how this unusual dynamics of a molecule 
 can be observed. One evident possibility is to detect radiation due 
 to excited dipole oscillations from a beam of molecules flying above 
 the grating. The density $n$ in a  molecular beam moving with 
 supersonic velocity in a supersonic jet or Laval nozzle can be as 
 high as $10^{17}$ cm$^{-3}$. However, since only the molecules 
 moving very close to the surface ($L \leq 0.1$ $\mu$m in our 
 numerical example) are efficiently excited, the total amount of 
 excited molecules above the grating plate of $1$ cm $\times 10$ 
 cm size will be of order $N \sim 10^{13}$. Each molecule 
 radiates with a power of order $W_1 \simeq \gamma \hbar \omega_0$. 
 At a frequency $\omega_0 = 10^{11}$ s$^{-1}$ their total spontaneous 
 emission power will be very small: $N W_1 \sim 10^{-16}$ W. To 
 increase the radiated power, one should pre-phase the dipole 
 oscillations of molecules at the entrance to the amplification 
 region by an external microwave field of intensity higher than the 
 thermal noise. Then the molecules will radiate coherently in the 
 bunches of length of the order of wawelength $\lambda$, and the power 
 radiated by one bunch will be $N^2$ times the power radiated by a 
 single molecule, where $N \sim n \lambda^2 R_0$.  For the above 
 values of $n$ and $\lambda \sim 1$ cm we obtain the total power in the microwatt range which 
 is easily detectable.

 Another way to study the dynamics of molecules is to observe the 
 change in absorption of a microwave radiation propagating through 
 the layer of parametrically excited molecules. This  
 scheme is also easily realizable.

 To increase further the radiated power and to facilitate the 
 requirements for experiments 
 it is desirable to launch molecules with higher speed. This 
 increases the frequency of parametric modulation and leads to 
 greater power of spontaneous emission (which goes as $\omega^4$). 
 Then one can excite vibrational transitions in molecules and  
 obtain an emitter in the far-IR range where very few sources exist. 
 Alternatively, for a given resonant frequency we can increase the 
 spatial period $L$ of the grating and the distance $R$ from the 
 interface. However, it  is difficult to accelerate neutral molecules 
 to velocities higher than 1 km/s. A way to overcome this difficulty 
 is to accelerate molecular ions and then neutralize them by charge 
 exchange. This can be achieved by putting a buffer gas cell or a 
 thin foil before the entrance to the excitation region with grating interface.

 We conclude that there is one more fundamental effect in the realm 
 of QED, namely, the parametric self-excitation of dipole 
 oscillations of a molecule moving close to a periodic grating. This 
 effect can be experimentaly observed in different ways  and can be 
 even employed for generation of coherent radiation in the far-IR to 
 microwave ranges.  

 {\it Acknowledgments.} We thank Marlan Scully for valuable 
 discussions and  appreciate the support from the Texas Advanced Technology Program, the  Office of Naval research, and High Energy Lasers Program of Department of Defense.


\newpage
 \begin{thebibliography}{}

\bibitem{Dr74} K.H.~Drexhage, in {\em Progress in Optics}, edited by
E.~Wolf, {\bf 12}, 163 (1974).

\bibitem{Ch78} R.R.~Chance, A.~Prock, and R.~Silbey, Adv. Chem. Phys.
{\bf 37}, 1 (1978); \\ J.M.~Wylie, J.E.~Sipe, Phys. Rev. {\bf A 30}, 
1185 (1984); Ibid. {\bf A 32}, 2030 (1985).

\bibitem{Hi91} E.A.~Hinds, Adv. Atom. Mol. Opt. Phys. {\bf 28}, 237
(1991); Adv. Atom. Mol. Opt. Phys. Suppl. {\bf 2}, 1 (1994).

\bibitem{Mo69} H.~Morawitz, Phys. Rev. {\bf 187}, 1792 (1969); \\ 
P.W.~Milonni, P.L.~Knight, Opt. Comm. {\bf 9}, 119 (1973).

\bibitem{Ar88} H.F.~Arnoldus, T.F.~George, Phys. Rev.  {\bf A 37}, 761
(1988).

\bibitem{Kh91} H.~Khosravi, R.~Loudon, Proc. Roy. Soc. Lond.  {\bf
A 433}, 337 (1991); Ibid. {\bf A 436}, 373 (1992); \\
S.M.~Barnett, B.~Huttner, and R.~Loudon, Phys. Rev. Lett. {\bf 68}, 3698 
(1992).

\bibitem{Or94} A.N. Oraevsky, Usp. Fiz. Nauk {\bf 164}, 415
(1994).

\bibitem{Ch95} V.F.~Cheltsov, Nuovo Cim. {\bf D 17}, 17 (1995).

\bibitem{Me92} D.~Meschede, Phys. Rep. {\bf 211}, 201 (1992).

\bibitem{Ya92} Y.~Yamamoto, S.~Machida, and G.~Bjork, Opt. Quant.
Electr. {\bf 24}, 215 (1992).


\bibitem{Ca48} H.~Casimir, D.~Polder, Phys. Rev. {\bf 73}, 360
(1948).

\bibitem{Ba70} G.~Barton, Proc. Roy. Soc. Lond. {\bf 320}, 251
(1970); Ibid. {\bf A 410}, 141, 175 (1987); J. Phys. {\bf B 7}, 2134
(1974).
\bibitem{Mi82} P.W.~Milonni, Phys. Rev.  {\bf A 25}, 1315 (1982);
Phys. Scripta {\bf T 21}, 102 (1988).

\bibitem{Da82} J.~Dalibard, J.~Dupont-Roc, and C.~Cohen-Tannoudji, J.
de Physique {\bf 43}, 1617 (1982); Ibid. {\bf 45}, 637 (1984).

\bibitem{Me90} D.~Meschede, W.~Jhe, and E.A.~Hinds, Phys. Rev. {\bf A
41}, 1587 (1990).

\bibitem{Hi91a} E.A.~Hinds, V.~Sandoghdar, Phys. Rev. {\bf A 43}, 398
(1991).

\bibitem{Ba87} A.O.~Barut, J.P.~Dowling, Phys. Rev. {\bf A 36}, 649
(1987).


\bibitem{Mi73} G.S.~Agarwal, Phys. Rev. {\bf A 12}, 1475 (1975); \\ 
G.S.~Agarwal, S.D.~Gupta, R.R.~Puri, {\em Fundamentals of cavity 
quantum electrodynamics}, World Scientific, Singapore (1995); \\ 
H.~Morawitz, M.R.~Philpott, Phys.  Rev.  {\bf B 10}, 4863 (1974).

\bibitem{By72} V.P.~Bykov, Zh. Eksp. Teor. Fiz. {\bf 62}, 505 (1972),
Ibid. {\bf 63}, 1226 (1972) (Sov. Phys. - JETP {\bf 36}, 646 (1973)); 
Kvant. Elektron. (Moscow) {\bf 1}, 1557 (1974) (Sov. J. Quant. Electron. 
{\bf 4}, 861 (1975)).

\bibitem{Sa94} S.~John, T.~Quang, Phys. Rev. {\bf A 50}, 1764
(1994).

\bibitem{Su95} T.~Suzuki, P.K.L.~Yu, J. Opt. Soc. Am. {\bf B 12}, 570
(1995); M.~Scalora, J.P.~Dowling, M.~Tossi, M.J.~Bloemer, 
C.M.~Bowden, J.W.~Haus, Appl. Phys. B, {\bf 60}, S57 (1995). .

\bibitem{bkk95}   A.A. Belyanin, 
 V. V. Kocharovsky, and Vl.V. Kocharovsky,
 Laser Physics {\bf 5},  1164 (1995). 
 

\bibitem{bkk96} 
 V.V. Kocharovsky,  Vl.V. Kocharovsky, and  A.A. Belyanin, 
  Phys. Rev.
Lett. {\bf 76}, 3285-3288 (1996).

 \bibitem{gin} V.L. Ginzburg and V.M. Fain, JETP {\bf 35}, 817 (1958). 

 \bibitem{sp} S.J. Smith and E.M. Purcell, Phys. Rev. {\bf 92}, 
 1069 (1953). 

\bibitem{tr} V.L. Ginzburg and I.M. Frank, JETP {\bf 16}, 15 (1946). 

\end{thebibliography}
\end{document}